\documentclass[showpacs,aps,prl,twocolumn,superscriptaddress]{revtex4-1}
\usepackage{graphicx} 
\usepackage{dcolumn}
\usepackage{bm}
\usepackage{amssymb,amsmath}
\usepackage{epstopdf}
\usepackage{color}
\def\etal{$\it{et~al.}$}
\newcommand{\fref}[1]{Fig.~\ref{#1}}

\begin{document}

\title{Effect of uniaxial strain on the optical Drude scattering in graphene}

\author{Manisha Chhikara}
\affiliation{Department of Quantum Matter Physics, University of Geneva, CH-1211 Geneva 4, Switzerland}
\author{Iaroslav Gaponenko}
\affiliation{Department of Quantum Matter Physics, University of Geneva, CH-1211 Geneva 4, Switzerland}
\author{Patrycja Paruch}
\affiliation{Department of Quantum Matter Physics, University of Geneva, CH-1211 Geneva 4, Switzerland}
\author{Alexey. B. Kuzmenko}
\email{Alexey.Kuzmenko@unige.ch}, \email{Manisha.Chhikara@unige.ch}
\affiliation{Department of Quantum Matter Physics, University of Geneva, CH-1211 Geneva 4, Switzerland}

\date{\today}

\begin{abstract}
\textbf{Graphene is a mechanically robust 2D material promising for flexible optoelectronic applications. However, its electromagnetic properties under strain are experimentally poorly understood. Here we present the far-infrared transmission spectra of large-area chemical-vapor deposited monolayer graphene on a polyethylene terephthalate substrate subjected to uniaxial strain. The effective strain value is calibrated using the Raman spectroscopy and corrected for a relaxation of wrinkles and folds seen directly by atomic-force microscopy. We find that while the Drude weight and the Fermi level remain constant, the scattering rate increases by more than 10\% per 1\% of applied strain, showing a high level of reproducibility during strain cycling. As a result, the electronic mobility and optical absorption of graphene at terahertz and lower frequencies appear to also be sensitive to strain, which opens pathways to control these key parameters mechanically. We suggest that such a functionality can be potentially used in flexible optoelectronic and microelectromechanical systems based on graphene. By combining our findings with existing theoretical models, we discuss the possible mechanisms of strain-controlled Drude scattering.}
\end{abstract}

\maketitle

\section{Introduction}

The remarkable electronic, optical and mechanical properties of graphene make it extremely promising for flexible optoelectronics \cite{BonaccorsoNP10,JangAM16}. In addition to its high conductivity and optical transparency, this two-dimensional material possesses exceptional mechanical flexibility with outstanding values of Young's modulus ($\sim$ 1 TPa) and tensile strength ($\sim$ 40 N/m) \cite{LeeScience08}, allowing it to sustain strains of up to 25\% \cite{LiuPRB07,KimNature09}.  The elastically driven deformation of the C-C bonds changes the vibrational and electronic spectra and therefore affects a large number of physical properties of graphene. Uniaxial strain, which is the easiest type of deformation to implement, was experimentally found to influence the electrical resistivity \cite{KimNature09,ShioyaNL15}, the work function \cite{HeAPL15}, Raman peaks \cite{YuJPCCL08,MohiuddinPRB09,CorroJPC15}, and visible light transmission spectra \cite{NiAM14}.

Much less is known about the effect of strain on the electromagnetic Drude absorption by free carriers, which determines the far-infrared, terahertz, and microwave properties of graphene, and is therefore essential for low-energy optoelectronic and plasmonic applications. Pellegrino \etal\  \cite{PellegrinoHPR10} calculated the strain dependence and the induced anisotropy of the Drude weight, based on the well established tight-binding theory of uniaxially strained graphene \cite{PereiraPRB09,RibeiroNJP09}. In the only experimental report known to us \cite{KimAPL12}, Kim \etal\  studied the strain evolution of the far-infrared response of graphene on a stretchable low-density polyethylene (LDPE) substrate under a very high amount of strain (15\%). Unexpectedly, the authors observed an incoherent non-Drude behavior for the polarization parallel to the strain axis. However, since the measurements were not reproducible during strain cycling, it remains unclear whether this observation is related to the intrinsic properties of strained graphene or is a result of structural damage. Achieving systematic and reproducible results in this case is not only a must to establish the physical mechanisms behind strain evolution, but also a key requirement for benchmarking flexible optoelectronic device applications. This motivated us to undertake a systematic and quantitative investigation of the effect of strain on the intrinsic Drude absorption in graphene.

Here we use Fourier transform infrared (FT--IR) spectroscopy as the primary technique to probe the Drude absorption and measure optical conductivity in chemical-vapor deposited (CVD) graphene on an elastic polyethylene terephthalate (PET) substrate under mechanical deformation. We achieve a high level of reproducibility of the measurements under consecutive strain sweeps. The graphene surface presents multiple folds and wrinkles generated during the transfer of graphene onto the substrate, which we characterize with nanoscale precision via atomic force microscopy (AFM). By bending the substrate, we apply a nominal tensile strain of up to 2\%. We observe directly that wrinkles flatten when the substrate is bent, which suggests that the actual strain may be lower than the value expected under the assumption that graphene precisely follows the substrate surface. Therefore, we use Raman spectroscopy to extract the true strain value by observing the shift in the 2D Raman peak. The effective strain is indeed found to be lower than the nominal value. Our results show a significant variation of the optical conductivity of graphene in the far-infrared regime, even under relatively small effective strain below 1\%. Importantly, the Drude weight remains constant within the experimental error bars, with the substrate apparently acting as a charge reservoir, while the scattering rate increases strongly with strain, most probably due to increased contact and stronger interactions between the graphene and the underlying substrate. This control of optical conductivity, potentially useful for flexible optoelectronic device applications, is highly reproducible. To complement the experimental investigation, we present detailed theoretical analysis of the effect of strain on the Drude absorption in graphene, involving various scattering mechanisms.

\section{Methods}
\subsection{Samples and strain application}

We used commercially available CVD-grown monolayer graphene (\emph{Graphenea}) transferred onto a 250 $\mu$m thick PET slab. This substrate is sufficiently transparent and flexible to allow precise and reproducible control of strain while doing optical and scanning-probe measurements. Large-area square samples (11$\times$11 mm$^{2}$) were used to ensure the uniformity of the strain in the middle of the sample and to avoid any diffraction effects at the lowest optical frequencies investigated.

\begin{figure*}
\includegraphics[width=15cm]{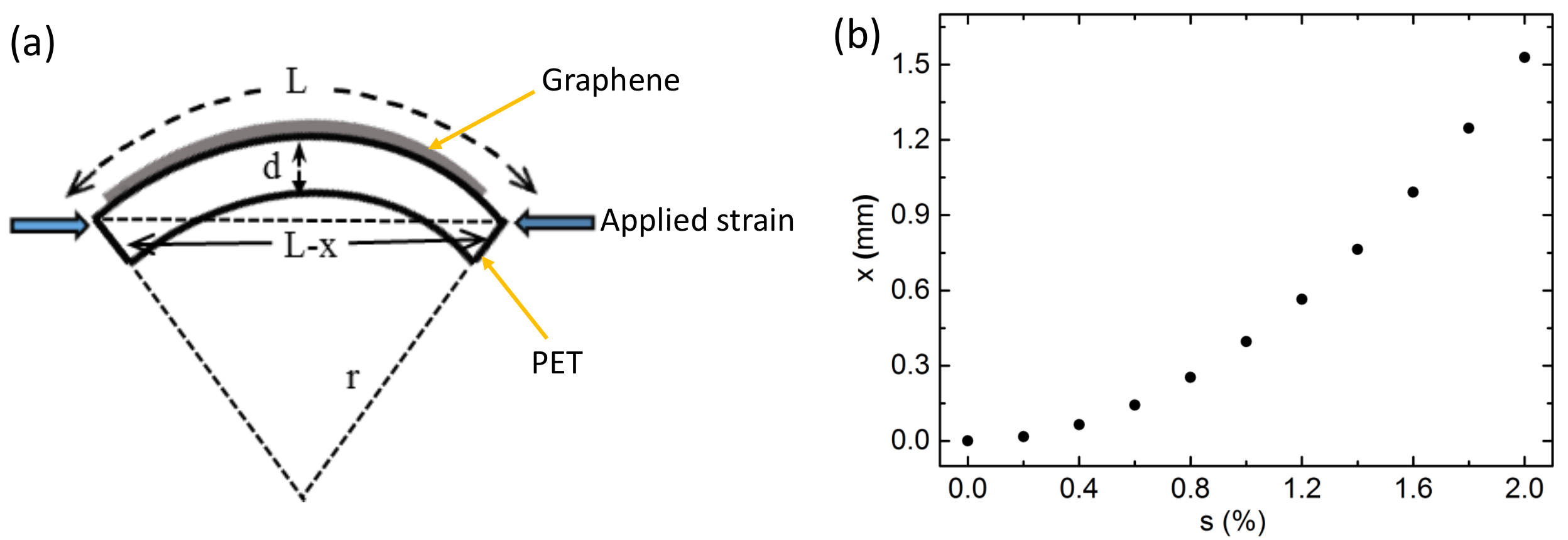}
\caption{(a) Schematic diagram of the application of tensile strain to CVD graphene by bending the underlying $L\times L$ PET substrate of thickness $d$. Bending is realised via a precise control of the shift parameter $x$, which determines the bending radius $r$. (b) The nominal strain $s=d/(2r)$ of graphene  as a function of $x$, assuming the absence of slipping.}
\label{strain_application}
\end{figure*}

Tensile strain was applied to the graphene by bending the substrate, with graphene on the convex side, using a home-made micrometer-based mechanical device, similar to the the method used in Refs. \cite{MohiuddinPRB09,NiAM14}, as shown in \fref{strain_application} and \fref{optical_setup}. The nominal applied strain (assuming that graphene follows the substrate without slipping) is calculated as $s = d/2r$, where $d$ is the substrate thickness substrate and $r$ is the bending radius. Assuming a uniform bending, the micrometer displacement $x$ and the strain are related by the formula $x=L[1-(d/sL)\sin(sL/d)]\approx L^3s^2/(6d^2)$, where $L$ is the substrate width, as shown in \fref{strain_application}(b). The same bending method was used in all measurements presented in this paper.

\subsection{Experimental methods}

The atomic force microscopy (AFM) and Kelvin probe force microscopy (KPFM) measurements were performed  with Ti/Ir coated tips (resonance frequency $\sim$ 70 KHz) under controlled humidity (30$\pm$3\%) at room temperature. These measurements proceed in two-pass scan mode: in the first pass, a normal topography height image is obtained in tapping mode; in the second pass, the cantilever is lifted 30 nm above the surface and follows the previously obtained topography to record the contact potential difference (CPD) between the graphene and the tip.

The Raman measurements were performed at the excitation wavelength of 514.5 nm (2.41 eV) with a 50$\times$ objective. The spot size was 1 $\mu$m and the laser power less than 2 mW to avoid laser-induced heating. Due to the dominance of PET-induced Raman peaks over the phonon G peak of graphene ($\sim$1580 cm$^{-1}$), we have focused our analysis on the shift of the double-resonance 2D peak ($\sim$2700 cm$^{-1}$), which is moreover very sensitive to strain \cite{YuJPCCL08,MohiuddinPRB09,CorroJPC15}.

\begin{figure*}
\centerline{\includegraphics[width=18cm]{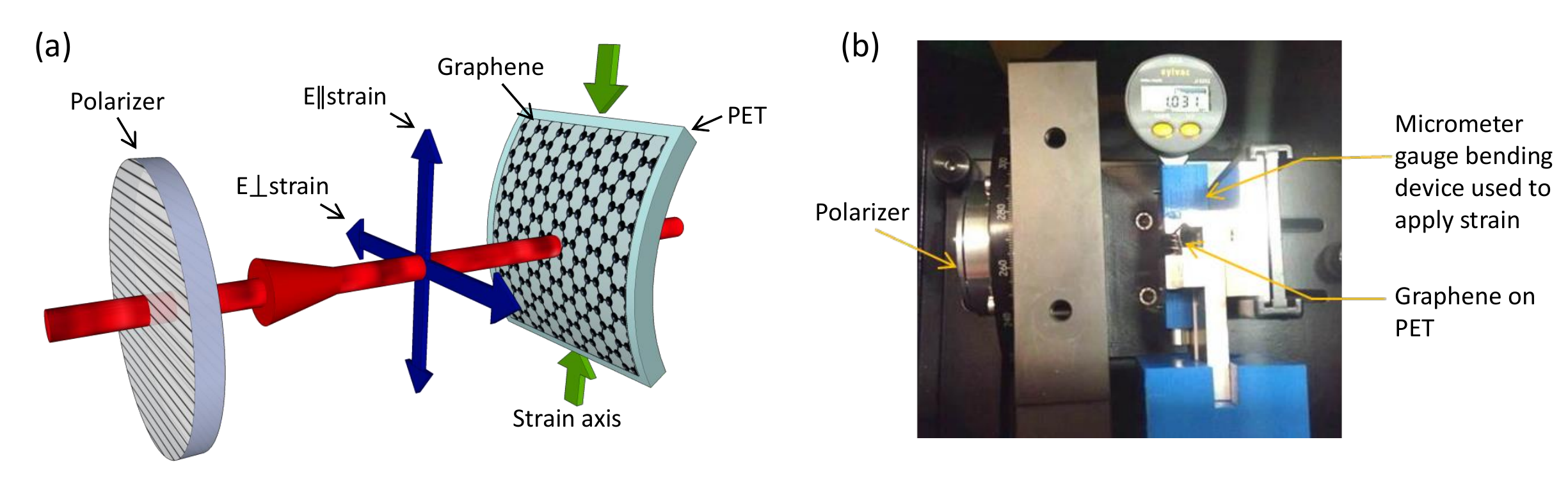}}
\caption{(a) Schematic diagram of the FT--IR experiment. (b) A photograph of the experimental setup including a polarizer and home-built micrometer-based device used to bend the substrate.}
\label{optical_setup}
\end{figure*}

Infrared transmission spectra were measured using a standard FT--IR spectrometer (\emph{Bruker Vertex 70v}) over the 40--700 cm$^{-1}$ frequency range, using a globar light source and a liquid-He cooled Si bolometric detector. A spot size of 2 mm was used to keep light well in the middle of the sample. Absolute transmission at every strain value was normalized against the empty sample holder at the same micrometer position $x$ in order to keep the measurement conditions exactly the same. Spectra for both graphene/PET samples, $T_{\text{g}}(\omega)$ and bare reference substrates $T_{\text{s}}(\omega)$ were obtained separately for incident light polarized parallel ($E_{\parallel}$) and perpendicular (E$_{\perp }$) to the strain direction as sketched in \fref{optical_setup}.

All measurements were performed at room temperature. They were repeated during several strain cycles and showed excellent reproducibility.

\section{Results and discussion}

\subsection{Scanning probe and Raman measurements}

\begin{figure*}
\includegraphics[width=12cm]{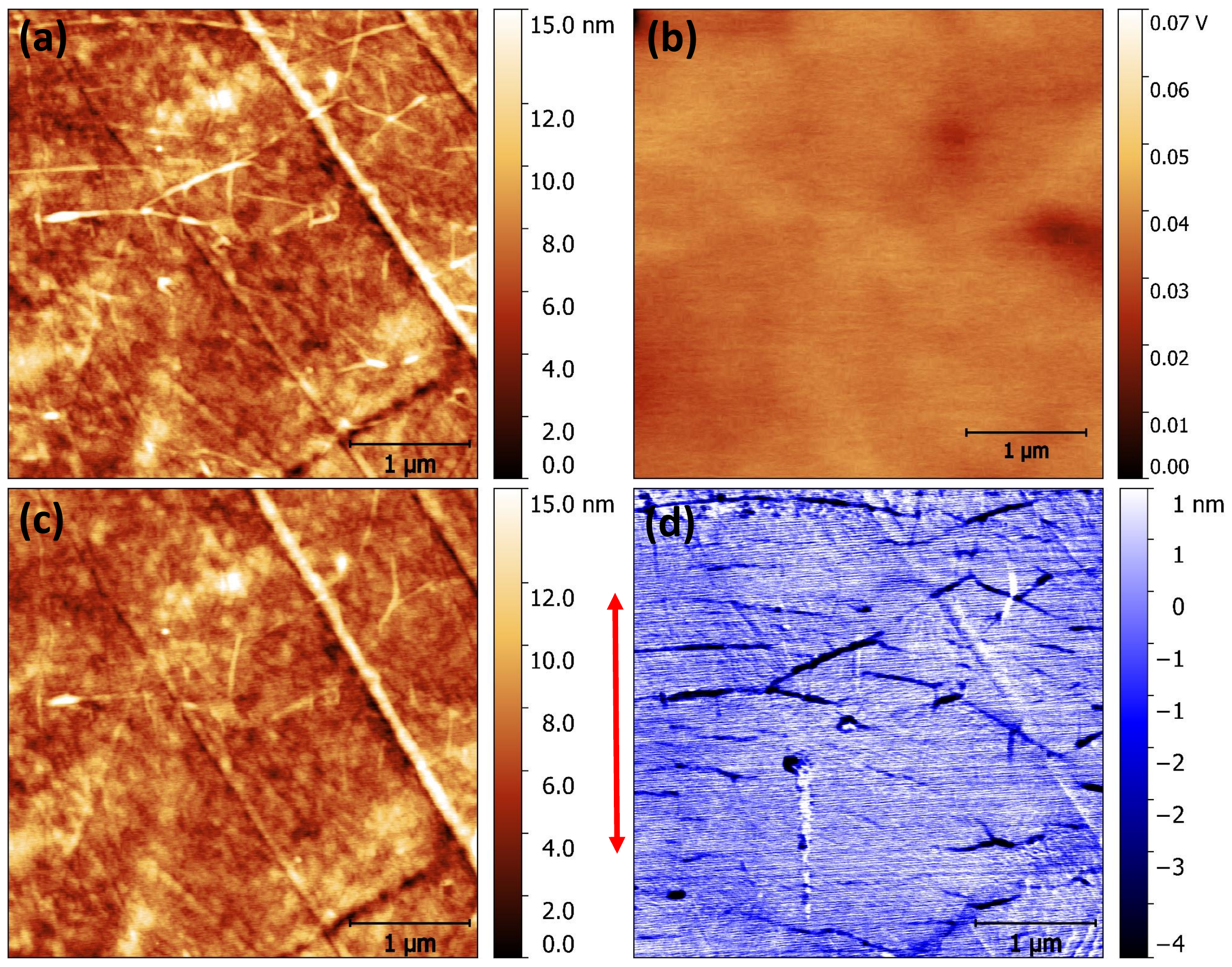}
\caption{4$\times$4 $\mu$m$^2$ scanned probe micrographs showing (a) the topography and (b) the contact potential difference (CPD) of unstrained graphene. Under 2\% nominal strain (c), the topography evolves , showing a relaxation of wrinkles and folds on the graphene surface. From a differential comparison (d) of the two topographies after correction of the images for scanner drift \cite{Gaponenko_ms}, this relaxation can be correlated with the axis along which the strain is applied to the graphene (red arrow).}
\label{graphene_AFM}
\end{figure*}

In \fref{graphene_AFM}(a) and (b) we show a typical a 4$\times$4 $\mu$m$^2$ topographical micrograph and contact potential difference map renormalized around the mean value of the same region of unstrained graphene. The topography, representative of measurements performed at multiple different locations on the sample, reveals a clean graphene surface almost free from chemical residues, and is characterized by wrinkles and folds of different heights and orientations. For the CPD, while absolute values could not be determined in the absence of a reference calibration, we note that only very small variations are observed across the graphene surface, with a root mean square ($V_{\text{rms}}$) value of 3.4 mV, which is limited by the resolution of KPFM setup at ambient conditions. When the same area is measured under higher nominal strain ($s$ =1.2, 1.6 and 2\%) no significant evolution of the $V_{\text{rms}}$ values is observed.  However, as can be seen in \fref{graphene_AFM}(c) for the topographical micrograph at 2\% nominal strain, a number of small folds and wrinkles in the graphene appear to have relaxed.  To quantify this relaxation behavior, we performed a differential analysis of the topographical images using an in-house developed drift correction algorithm \cite{Gaponenko_ms}, which allows changes from one scan to another of the same area to be determined with sub-nm precision. \fref{graphene_AFM}(d) shows the resulting differential image, with the dark contrast features corresponding to all the wrinkles and folds that relaxed after the application of 2\% nominal strain. They show decreased height with respect to the unstrained graphene imaged in \fref{graphene_AFM}(a).  We can see that relaxation occurs primarily for wrinkles aligned perpendicular to the axis of applied strain, indicated by the red arrow, as would be logically expected. The only clear increase in the height of a fold, occurring in the star-like feature at the upper right of the image, was observed for a wrinkle aligned parallel to the the axis of applied strain, and accompanied relaxation in the rest of the feature.  These strong variations of the nanoscale topographic landscape through relaxation under the applied strain imply that the effective strain transferred to the sample might be significantly smaller than the nominal strain.

\begin{figure*}
\includegraphics[width=15cm]{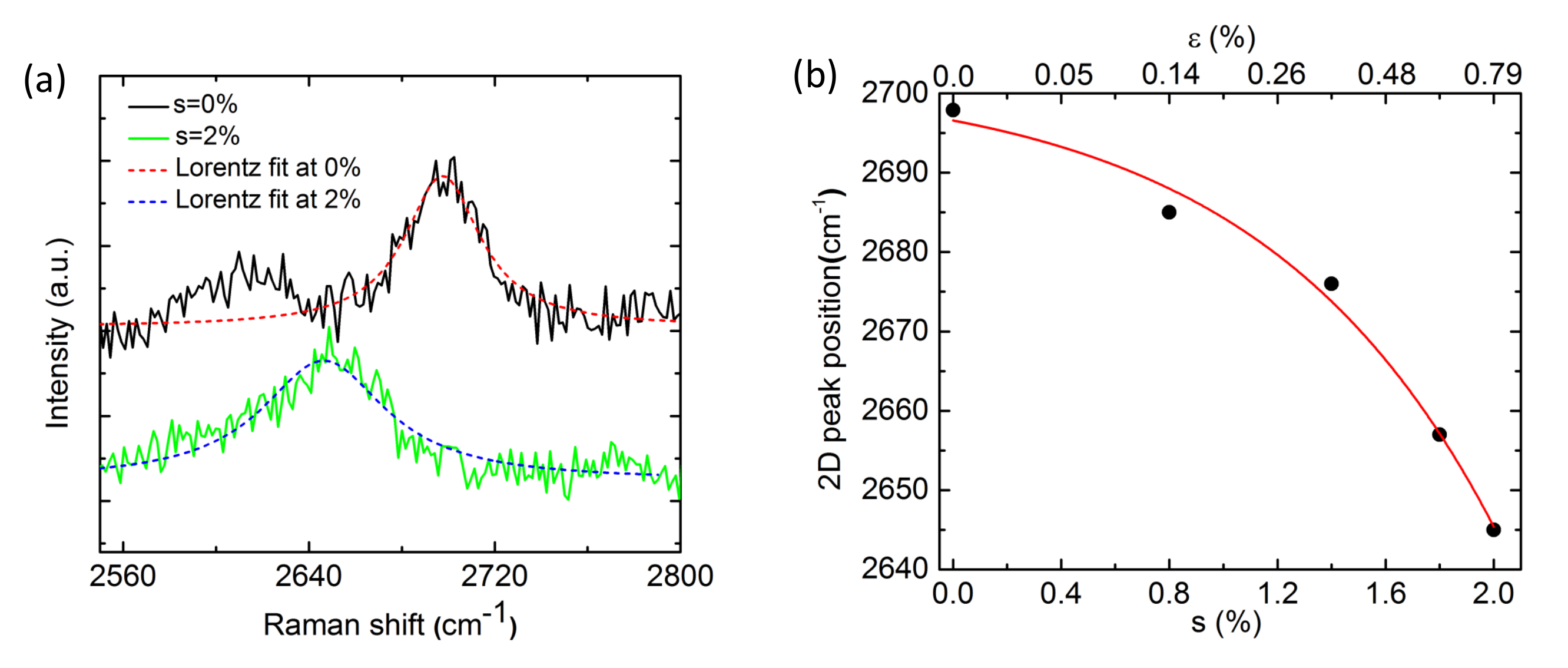}
\caption{(a) Raman spectra of graphene under 0\% and 2\% of nominal strain, together with their Lorentzian fits, showing a red shift of the 2D peak. (b) 2D peak position as a function of the nominal strain $s$, allowing the effective strain $\epsilon$ to be obtained by calibration to measurements on exfoliated graphene \cite{MohiuddinPRB09}.}
\label{Raman}
\end{figure*}

The actual amount of strain applied to graphene was therefore calibrated with Raman measurements. At 2\% nominal strain, we observed a 52 cm$^{-1}$ red shift caused by the elongation and weakening of C-C bonds, displacing the 2D peak from 2697 to 2645 cm$^{-1}$ as shown in \fref{Raman}(a). The solid black and green lines show the experimental data measured at 0\% and 2\% nominal strain, respectively, fitted with a Lorentzian function (red dashed line for 0\% and blue dashed line for 2\%). The observed red shift in the position of the 2D peak is smaller than the values of 64 cm$^{-1}$ per \% uniaxial strain reported by Mohiuddin \etal, on a mechanically exfoliated single layer graphene flake free of wrinkles and grain boundaries associated with CVD-grown material \cite{MohiuddinPRB09}. In our case, we must therefore presume that the observed relaxation of folds/wrinkles leads to a sliding at the graphene-PET interface \cite{HeAPL15}, and use the single-grain data as a reference to calibrate the effective vs. nominal strain in our measurements. \fref{Raman}(b) shows the  2D peak position (black circles) as a function of nominal strain $s$ (bottom abscissa) and the corresponding effective strain $\epsilon$ (upper abscissa). The corresponding values of the red shift $\Delta\omega_{2D}$ were extracted from a fit of the data (red line). $\epsilon$ is then obtained using the relation $\Delta\omega_{2D} / \epsilon = 64$ cm$^{-1}$ per \% uniaxial strain, based on the exfoliated graphene results \cite{MohiuddinPRB09}.

\subsection{Far-infrared measurements}

\begin{figure*}
\includegraphics[width=17cm]{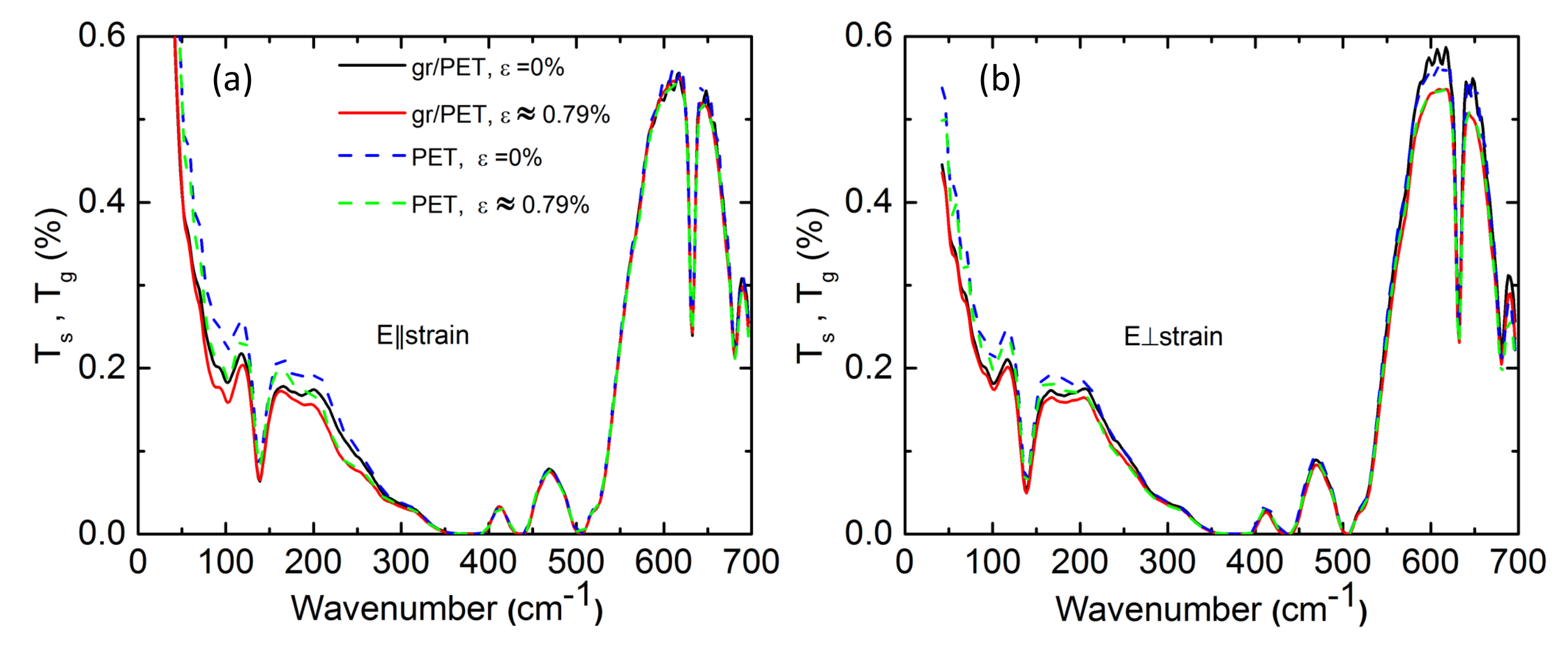}
\caption{Absolute far-infrared transmission spectra through the bare PET substrate (dashed lines), $T_{\text{s}}(\omega)$, and through graphene on the substrate (solid lines), $T_{\text{g}}(\omega)$, for $\epsilon$= 0\% and 0.79\% (the maximum effective strain) . Panels (a) and (b) show results for the polarization parallel and perpendicular to the strain axis.}
\label{FigTran}
\end{figure*}

As the PET substrate demonstrates strong photoelastic effects (as discussed in the Supplementary Information) it is necessary to fully characterize it optically as a function of the bending radius before the extraction of the intrinsic graphene properties. To this end we used a reference graphene-free substrate of the identical thickness, taken from the same batch. In \fref{FigTran} the absolute far-infrared transmission spectra $T_{\text{s}}(\omega)$ are presented for the polarization of light parallel (a) and perpendicular (b) to the strain axis. The spectra for a perfectly flat and a maximally bent substrate substrate, corresponding to the effective graphene strain of 0.79\%, are shown by the dashed blue and green lines, respectively. One can see that the substrate is sufficiently transparent everywhere except in the region between 300 cm$^{-1}$ and 540 cm$^{-1}$, where strong phonon absorption is present. The observed minima at 138 cm$^{-1}$ and 632 cm$^{-1}$ are due to weaker optical phonons in PET, which do not fully block the transmission. Periodic oscillations due to the Fabry-Perot interference in the substrate are seen, as expected, in the spectral regions of high transparency. For both polarizations, applying strain results in a frequency-dependent reduction of the transmission due to the above-mentioned photoelastic effects. Applying a Kramers-Kronig analysis \cite{KuzmenkoRSI05} to these data allows us to extract the spectra of the complex refractive index $N=\sqrt{\epsilon_{\text{PET}}}$  for each polarization, at each value of the applied strain, as detailed in the Supplementary information.

The solid black and red solid curves in \fref{FigTran} represent the absolute transmission $T_{\text{g}}(\omega)$ of graphene on PET for zero and the highest effective strain (0.79\%), respectively. The presence of the graphene monolayer noticeably reduces the transmission at low frequencies (below 200--300 cm$^{-1}$) with respect to the bare substrate, while it does not essentially affect the transmission at high frequencies. Qualitatively, this indicates that the transmission reduction is due to the absorption by Drude carriers in graphene, as observed previously \cite{HorngPRB11,KimAPL12,SensaleNatComm12}.

\begin{figure*}
\includegraphics[width=18cm]{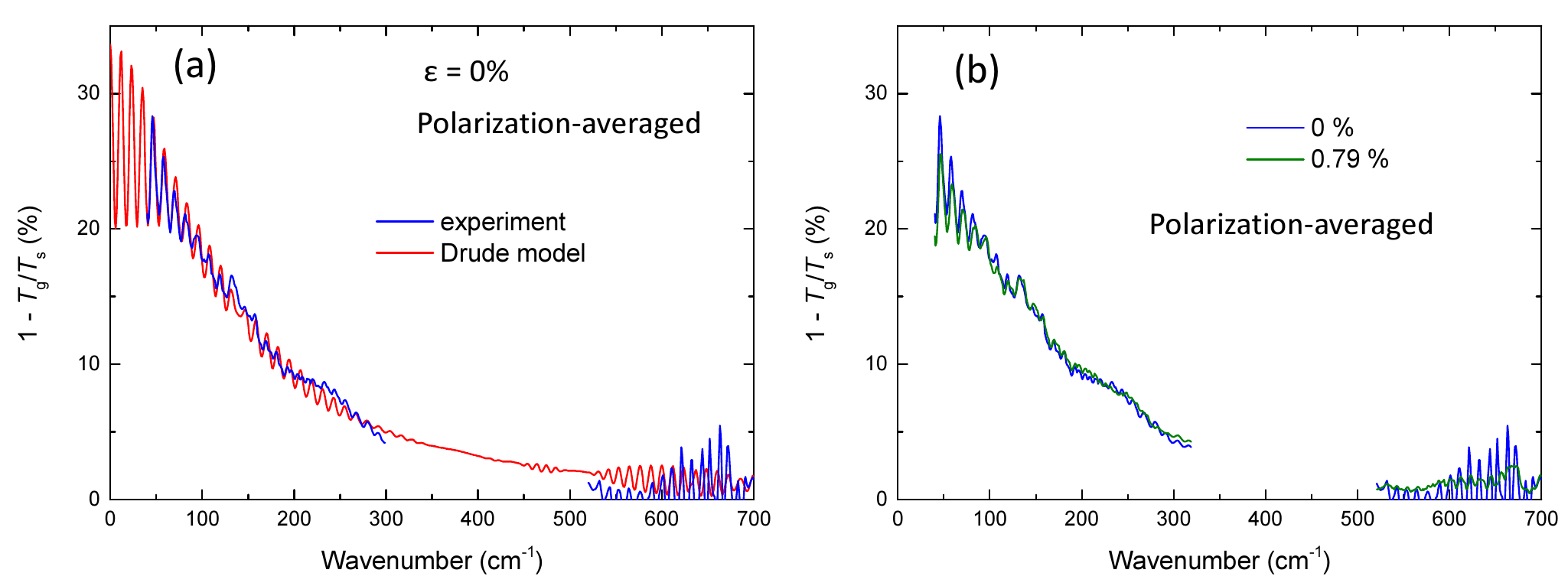}
\caption{(a) Experimental polarization-averaged extinction spectrum of graphene at zero strain (blue curve), together with a Drude fit (red curve) using the independently calibrated optical functions of PET. (b) Experimental extinction spectra at zero strain (blue) and at the maximum effective strain $\epsilon=$ 0.79\% (green).}
\label{FigExtModel}
\end{figure*}

In order to see the Drude absorption more clearly, in \fref{FigExtModel} we plot the extinction spectra $1-T_{\text{g}}(\omega)/T_{\text{s}}(\omega)$, where the effects of the substrate are largely suppressed as compared to the absolute transmission. Furthermore, in order to avoid any spurious effects originating from a possible light depolarization in the substrate, from this point on we discuss the polarization-averaged, instead of polarization-resolved, spectra. \fref{FigExtModel}(a) shows the extinction at zero strain (solid blue line), where a pronounced Drude peak is seen, with the maximum extinction reaching about 25\% at the lowest frequencies. For an ultrathin film on a substrate the extinction is given by:
\begin{equation}\label{TgTs}
1-\frac{T_{\text{g}}}{T_{\text{s}}}=1-\frac{1}{\left| 1+f_{\text{s}} Z_{0}\sigma/2\right|^{2}},
\end{equation}
\noindent where $Z_{0} =4\pi/c\approx $ 377 $\Omega$ is the vacuum impedance, $\sigma=\sigma_{1}+i\sigma_{2}$ is the two-dimensional optical conductivity of graphene and
\begin{equation}\label{TgTs}
f_{\text{s}}=2\frac{(N+1)+(N-1)\exp(2i\omega N d/c)}{(N+1)^2-(N-1)^2\exp(2i\omega N d/c)}.
\end{equation}
\noindent The substrate factor $f_{\text{s}}$ reduces the extinction as compared to the case of free standing graphene, where it is equal to one, and adds some Fabry-Perot oscillations, clearly seen in the experiment. The Drude model for the optical conductivity reads as \cite{HorngPRB11}:
\begin{equation}\label{Drude}
\sigma(\omega)=\frac{D}{\pi}\frac{\tau}{1-i\omega\tau},
\end{equation}
\noindent where $D$ is the Drude weight and $\tau$ is the scattering time. The red curve in \fref{FigExtModel}(a) presents the best fit using the Eqs (\ref{TgTs}) and (\ref{Drude}). One can see that overall the model works well: apart from perfectly reproducing the shape of the Drude peak, it also generates Fabry-Perot oscillations similar to the experimental curve. The amplitude of the oscillations is difficult to match using the model due to various dephasing effects in the substrate. Nevertheless, from this fit we can rather accurately extract the Drude parameters $D=1.39\pm 0.015$ k$\Omega^{-1}$ps$^{-1}$, and $\tau=0.28\pm0.005$ ps. We emphasize that the ability to measure $D$ and $\tau$ independently is one of the advantages of the infrared spectroscopy as compared to the DC transport measurements. Within the theory of non-interacting Dirac fermions \cite{AndoJPSJ02}, the Drude weight is determined solely by the Fermi level $E_{\text{F}}$ with respect to the Dirac point: $D=(e^2/\hbar^2)E_{\text{F}}$, where $\hbar$ is the reduced Planck constant and $e$ is the elementary charge. Therefore, this measurement allows us to determine the Fermi level $E_{\text{F}}=\hbar^{2}D/e^2 = 231\pm 2$ meV and the scattering rate $\hbar/\tau = 14.8 \pm 0.2$ meV. By adopting a standard value for the Fermi velocity $v\approx10^6$ ms$^{-1}$, we can estimate the electronic mobility $\mu = e v^{2}\tau/E_{\text{F}}\approx 2000$ cm$^2$V$^{-1}$s$^{-1}$ and the carrier concentration $n=E_{\text{F}}^{2}/(\pi\hbar^2 v^{2})\approx 4\times 10^{12}$ cm$^{-2}$. These quantities are consistent with typical values found in CVD graphene under ambient conditions  \cite{KimAPL12,HorngPRB11,RenNL12}.

The effect of strain on the extinction spectra is shown in \fref{FigExtModel}(b), where the blue and green curves correspond to $\epsilon$=0\% and 0.79\%, respectively. One can see that the Drude peak somewhat flattens under strain, showing a smaller (higher) value below (above) $\sim$ 120 cm$^{-1}$. This indicates that the scattering rate, which determines the width of the Drude peak, increases with strain, although a detailed analysis is required to quantify this effect, as carried out below. One can also notice that the amplitude of the Fabry-Perot oscillations at high frequencies decreases with strain, which is likely due to the substrate bending, not included in the flat-multilayer model, which may suppress the phase coherence between internally reflected rays.


\begin{figure*}
\includegraphics[width=18cm]{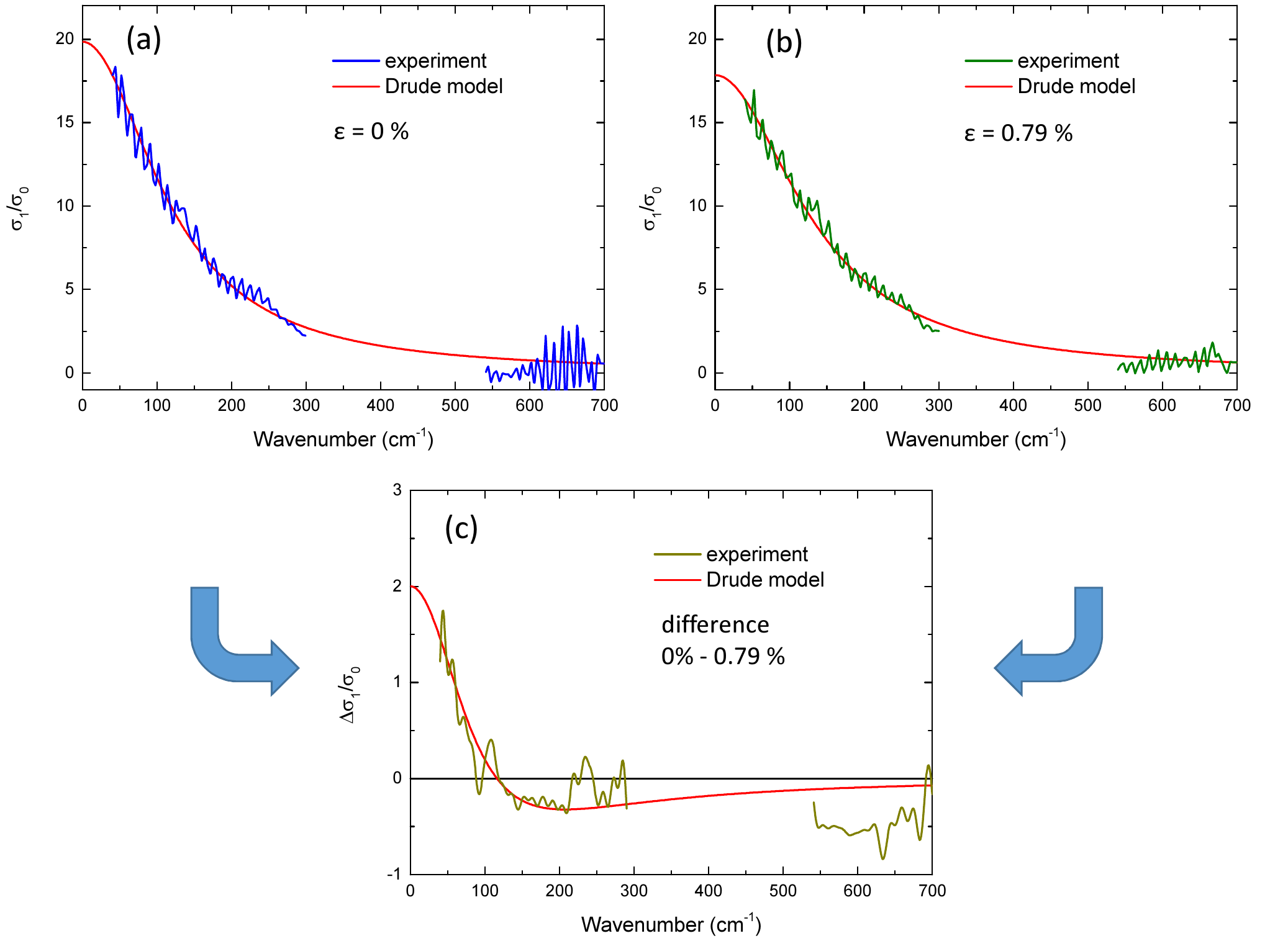}
\caption{Real part of the optical conductivity obtained by a Kramers-Kronig analysis \cite{KuzmenkoRSI05} normalized to the universal value $\sigma_{0}=e^2/(4\hbar)$. The data are shown for zero strain (a) and the maximum effective strain of 0.79\% (b) with the best Drude fits superimposed. Small oscillations are due to an incomplete compensation of the Fabry-Perot oscillations by the fitting procedure.  (c) The difference between the spectra at the two strain values , where the remnant Fabry-Perot oscillations are removed by Fourier filtering.}
\label{FigCond}
\end{figure*}

Before proceeding with a detailed analysis of the effect of strain, it is convenient to convert the extinction spectra into the optical conductivity of graphene using a model-independent Kramers-Kronig constrained analysis \cite{KuzmenkoRSI05}. \fref{FigCond}(a) shows the extracted spectrum (blue line) as well as the best Drude fit (red line) of the real part of the optical conductivity, $\sigma_{1}(\omega)$  at zero strain, normalized to the universal value $\sigma_{0} = e^{2}/4\hbar$ \cite{KuzmenkoPRL08,NairScience08}. The model matches the experiment well, apart from some remnant Fabry-Perot oscillations in the experimental curve, which appear due to the already-mentioned difficulty in precisely reproducing the oscillations in the extinction spectra. As one can see from \fref{FigCond}(b), the Drude model works equally well at the maximum strain. The zero-frequency limit of the optical conductivity $\sigma_{\text{dc}}=D\tau/\pi$ (obtained from the fit) shows a reduction from 20 $\sigma_{0}$ to 18 $\sigma_{0}$, \emph{i.e. } by about twice the total optical conductivity of graphene in the infrared and visible ranges \cite{KuzmenkoPRL08,NairScience08}. Thus the measured effect of strain on the optical conductivity in the far infrared regime is about two orders of magnitude larger than in the visible range \cite{NiAM14}. In order to emphasise the effect of strain, in \fref{FigCond}(c) we plot the experimental frequency-dependent differential conductivity $\Delta \sigma_{1} = \sigma_{1} (0\%)-\sigma_{1} (0.79\%) $ (dark yellow curve) and the corresponding difference between the Drude fits (red curve), where the remnant Fabry-Perot oscllations are Fourier-filtered. In spite of the unavoidable noise in the differential spectra, the match between experiment and the Drude model is obvious and a sign inversion is clearly observed at about 120 cm$^{-1}$. This fully justifies our preliminary conclusion that the Drude peak broadens under strain.


\begin{figure*}
\includegraphics[width=18cm]{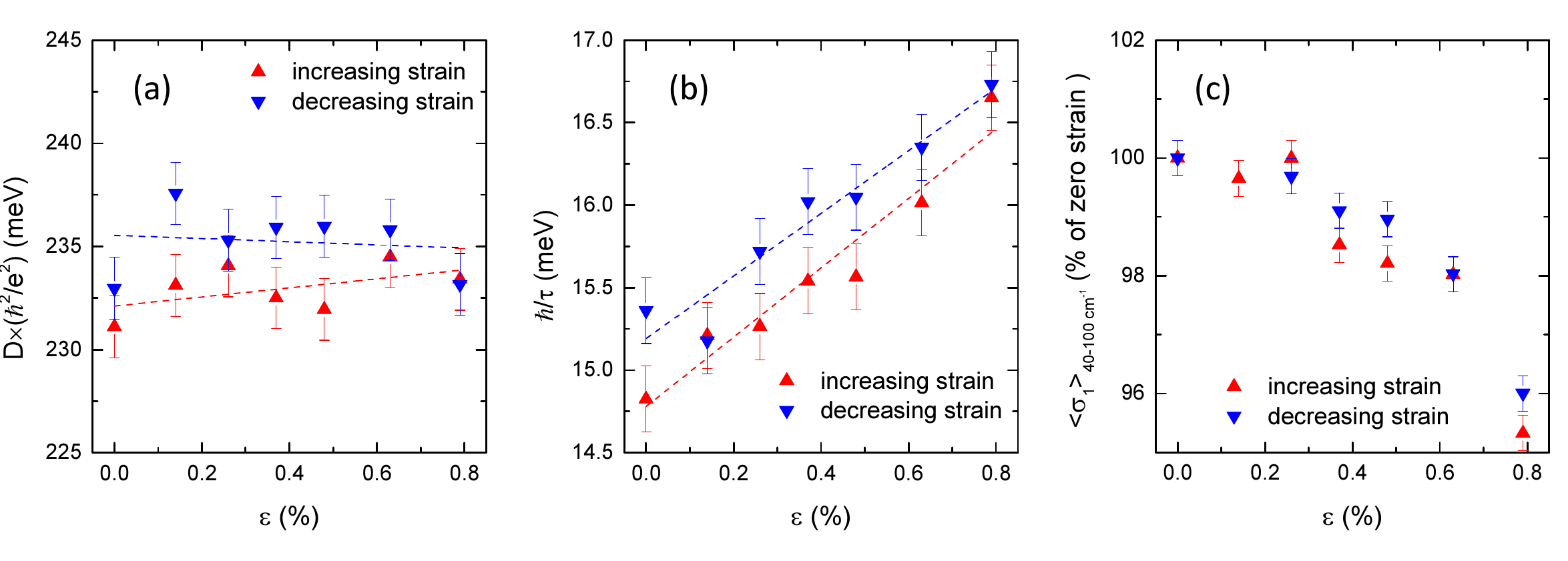}
\caption{Polarization-averaged Drude weight (a) and scattering rate (b) as a function of the effective strain extracted by fitting the extinction curve using the Drude model. The dashed lines are the best linear fits. (c) Optical conductivity integrated over the low-frequency interval 40-100 cm$^{-1}$ and normalized to zero-strain. In all panels, the data for increasing and decreasing parts of the strain cycle are shown by red and blue symbols respectively. }
\label{FigStrainDep}
\end{figure*}

 Using the same procedure, we extracted the Drude parameters independently at each value of strain. Furthermore, this was done separately for measurements carried out while increasing and decreasing the applied strain. Solid symbols in \fref{FigStrainDep}(a) and \fref{FigStrainDep}(b) show the Drude weight $\hbar^{2}D/e^2$ and the scattering rate $\hbar/\tau$, respectively, as a function of the effective strain for both phases of the cycle (red and blue). The dashed lines of the corresponding color are the linear fits. The central observation of this paper is that the Drude weight remains constant within the experimental error bars, while the scattering rate increases significantly with $\epsilon$, by more than 10\% at only 0.79\% of strain. It is important to note that the behavior of both the Drude weight and the scattering rate is reproducible and reversible.

\fref{FigStrainDep}(c) shows the measured value of the optical conductivity, $\sigma_{1}$, normalized to zero strain and spectrally averaged between 40 and 100 cm$^{-1}$ (1.2 - 3 THz), \emph{i.e.} the part of the experimental range where the optical conductivity decreases most significantly with $\epsilon$. One can see that the optical conductivity decreases reversibly by about 5\% for the maximum applied strain, which is driven in the present case, as we have shown, by the strain dependent Drude scattering rather than the Drude weight.

\subsection{Discussion of the Drude weight}

\begin{figure*}
\includegraphics[width=12cm]{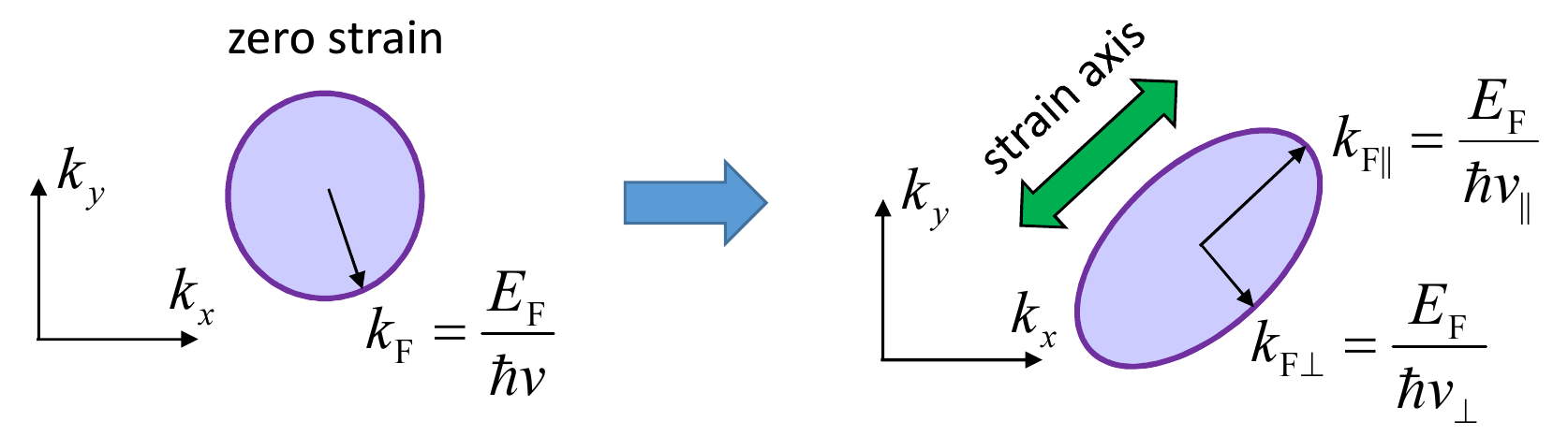}
\caption{Schematic representation of the deformation of the Fermi surface under uniaxial strain.}
\label{FigFermi}
\end{figure*}

The effect of uniaxial strain on the crystal lattice and the electronic band structure of graphene was extensively studied theoretically \cite{LiuPRB07,PereiraPRB09,RibeiroNJP09,PellegrinoPRB10,PereiraEPL10,BaimovaPSS12}. The nearest neighbor hopping parameter $t$ for the $\pi$-bands depends strongly on the bond distance $a$: $t/t_{0}\approx \exp[-3.37({a/a_{0}}-1)]$, where $t_{0}\approx $ 3 eV and $a_{0}\approx 0.14$ nm are the equilibrium values. Therefore, the major consequence of the lattice deformation  is to modify the three effective hopping parameters and subsequently to shift and uniaxially deform each of the six Dirac cones located in the K-points of the Brillouin zone. This gives rise to elliptical Fermi pockets elongated along the strain axis, regardless of the direction of the strain with respect to the crystal axes \cite{PellegrinoPRB10,PereiraEPL10} (Fig.\ref{FigFermi}). Thus, a uniaxial deformation is theoretically expected to anisotropically renormalize the Fermi velocity, producing two different values parallel ($v_{\parallel}$) and perpendicular ($v_{\perp}$) to the strain, with $v_{\parallel}<v_{\perp}$. This also modifies the density of states at the Fermi level:
\begin{equation}\label{DOS}
g(E_{\text{F}}) = \frac{2E_{\text{F}}}{\pi \hbar^2 v_{\parallel}v_{\perp}}
\end{equation}
\noindent and makes the Drude weight anisotropic \cite{PellegrinoHPR10}:
\begin{equation}\label{Daniso}
D_{\parallel, \perp} = \frac{\pi e^2g(E_{\text{F}})}{2}v_{\parallel, \perp}^2.
\end{equation}


Under small deformations, as in our experiment, we can apply a linear expansion for the strain-dependent Fermi energy:
\begin{equation}\label{EFLinear}
\frac{E_{\text{F}}}{E_{F,0}} \approx 1+\kappa \epsilon,
\end{equation}
\noindent and the (anisotropic) Fermi velocity:
\begin{equation}\label{VLinear}
\frac{v_{\parallel,\perp}}{v_{0}} \approx 1+\nu_{\parallel,\perp}\epsilon,
\end{equation}
\noindent where the index 0 refers to zero strain. As we show in the Supplementary Information, based on the existing \emph{ab-initio} calculations \cite{RibeiroNJP09}, the Fermi-velocity expansion coefficients are equal to $\nu_{\parallel} \approx -1.8$ and $\nu_{\perp} \approx +0.5$ and are independent of the strain orientation with respect to the crystal axis. This allows us to ignore the random orientation of the crystalline grains present in CVD graphene.

Using Equations (\ref{DOS})-(\ref{VLinear}), we can obtain a linear expansion for the density of states:
\begin{equation}\label{DOSLinear}
\frac{g(E_{\text{F}})}{g_{0}(E_{F,0})} \approx  1+( \kappa-\nu_{\parallel}-\nu_{\perp})\epsilon
\end{equation}
and the (anisotropic) Drude weight:
\begin{equation}\label{DanisoLinear}
\frac{D_{\parallel, \perp}}{D_{0}} \approx 1 + ( \kappa\pm\nu_{\parallel}\mp\nu_{\perp})\epsilon.
\end{equation}
\noindent One can see that $D_{\parallel}$ decreases, while $D_{\perp}$ increases with strain, which generates a Drude-weight dichroism $(D_{\parallel}-D_{\perp})/D_{0} \approx 2(\nu_{\parallel}-\nu_{\perp})\epsilon\approx -4.6 \epsilon$ determined solely by the anisotropy of the Fermi velocity and insensitive to the Fermi level. Measuring the anisotropy of the Drude weight would therefore directly probe the anisotropy of the Fermi velocity. Unfortunately, the photoelastic effects in the substrate did not allow us to measure the intrinsic dichroism of graphene with a sufficient precision, as discussed before. On the other hand, the average Drude weight $D=(D_{\parallel}+D_{\perp})/2\approx D_{0}(1+\kappa \epsilon)$ is measured very accurately, since the substrate depolarization does not add to the total optical absorption. This value depends only on the shift of the Fermi energy and is not affected by the anisotropy. Therefore, our observation of a constant (within error bars) Drude weight (\fref{FigStrainDep}(a)) indicates that the Fermi level is essentially strain-independent: $|\kappa| < 0.1$.

We note that this observation is a rather non-trivial one. Indeed, as the density of states varies with strain \cite{PereiraPRB09,PellegrinoPRB10,ShahMPLB13,HeAPL15}, one has to assume a strain-induced charge transfer between graphene and the environment (in particular, the substrate) in order to maintain $E_{\text{F}}$ constant. It is therefore likely that the PET substrate acts as a charge reservoir maintaining the chemical potential in graphene at a fixed level. Interestingly, our KPFM experiments show a rather small spatial variation of the chemical potential (about 3.4 meV), which may be also due to the stabilizing effect of the substrate on the chemical potential.

\subsection{Discussion of the strain-dependent scattering rate}

Next we discuss the surprisingly strong strain-induced increase of the scattering rate shown in \fref{FigStrainDep}(b). The data are well described by a linear dependence (dashed lines):
\begin{equation}\label{ScatLinear}
\frac{\hbar/\tau}{(\hbar/\tau)_{0}} \approx 1+c\epsilon,
\end{equation}
\noindent where the values $c=12.5 \pm 1$ and $14.2 \pm 1$ are found for the increasing and decreasing strain respectively. By taking the average value, we obtain the linear expansion coefficient $c =13.4 \pm 0.7$. Speaking differently, the scattering rate increases by about 13.5 \% per 1\% of the applied strain, which indicates that the Drude scattering is highly sensitive to the mechanical deformation.

In a recent theoretical report \cite{ShahMPLB13}, the effect of strain-dependent density of states on the scattering rate was calculated in a connection to transport measurements, and a significant increase of $\hbar/\tau$ with increasing strain was found. While the predicted effect was characterized as giant, the results of Ref. \cite{ShahMPLB13} are in fact smaller than what we observe here.

As it is widely discussed in the literature \cite{AndoJPSJ06,TanPRL07,HwangPRL07,StauberPRB07,ChenNP08,HwangPRB08,PonomarenkoPRL09,HwangPRB13,SongSR12,ShahMPLB13,YuPRB16}, the charge scattering in graphene is affected by short-range point defects (pd), long-range charged impurities (ci),  acoustic phonons (ap), surface-phonons (sp) in the substrate, and the graphene grain boundaries. According to the Matthiessen's rule, the total scattering rate is a sum of these respective contributions: $\hbar/\tau = \sum_{i} \hbar/\tau_{i} $. In principle, each of them may contribute to the total strain dependence. We can therefore attribute to each scattering channel ($i$) its respective strain-induced increase rate $c_{i}$:
\begin{equation}\label{ScatterGen}
c_{i} = \frac{\partial \ln(\hbar/\tau_{i})} {\partial \epsilon}.
\end{equation}
\noindent Even though we cannot distinguish the different scattering-rate contributions experimentally, it is clear that $c \leq \max(c_{i})$, \emph{i.e.} there should be at least one scattering channel, for which $c_{i}\gtrsim 13.4$. In the following discussion, we therefore consider each mechanism separately in order to identify candidate channels responsible for the large effect observed here. As we analyze the polarization-averaged spectra, and in order to use the existing theoretical literature, where only the isotropic case was considered, we neglect the strain-induce anisotropy of the Fermi surface and assume that the density of states changes with strain according to Eq. (\ref{DOSLinear}), where we furthermore set $\kappa = 0$, in agreement with our experiment.

\emph{Point defects. } Point defects are assumed to be uncharged short-range scatterers with a potential $V(\vec{r})=V_{\text{pd}}\delta(\vec{r})$ and the concentration $n_{\text{pd}}$ located directly in the graphene plane. They contribute to the scattering rate as follows \cite{HwangPRL07,StauberPRB07}:
\begin{equation}\label{ScatterPD}
\frac{\hbar}{\tau_{\text{pd}}} = \frac{\pi}{4} n_{\text{pd}} g(E_{\text{F}})V_{\text{pd}}^2.
\end{equation}
\noindent As the density of defects per unit cell remains the same, the spatial density changes inversely with the area of the unit cell, leading to $n_{\text{pd}}/n_{\text{pd},0}\approx 1-\epsilon$ , where we ignore a small Poisson ratio (0.16 in graphite \cite{BlaksleeJAP70,PereiraPRB09} but possibly even smaller in graphene on a substrate). Assuming that $V_{\text{pd}}$ does not change, we obtain:
\begin{equation}\label{ScatterPDLinear}
c_{\text{pd}}\approx -1 -2\nu \approx 0.3,
\end{equation}
\noindent where $\nu=(\nu_{\parallel} +\nu_{\perp} )/2\approx - 0.65$ is the average Fermi-velocity expansion coefficient. Such a small theoretical value of $c_{\text{pd}}$ allows us to safely exclude point defects as a candidate mechanism.

\emph{Charged impurities.} The scattering potential of charged impurities is given by: $V(q)=(2\pi e^2/\tilde{\kappa}q)e^{-qz}$, where $\tilde{\kappa}=(\epsilon_{\text{PET}}+1)/2\approx 1.8$ is the average dielectric constant of the media surrounding the graphene and $z$ is the distance between the layer containing these impurities (typically the substrate) and the graphene sheet \cite{HwangPRL07,StauberPRB07}. The contribution of charged impurities to scattering, which takes into account a dynamical self-screening by the graphene electrons, can be expressed as follows \cite{AndoJPSJ06,HwangPRL07,StauberPRB07} (see the Supplementary Information):
\begin{equation}\label{ScatterCI}
\frac{\hbar}{\tau_{\text{ci}}} = \frac{n_{\text{ci}}}{g(E_{\text{F}})}\int_{0}^{\pi}\frac{e^{-4k_{\text{F}}z\sin\frac{\theta}{2}}\sin^2\theta d\theta}{\left[1-\left(\frac{\pi}{4}-\frac{1}{2\alpha_{\text{g}}}\right)\sin\frac{\theta}{2}\right]^2}.
\end{equation}
\noindent Here $n_{\text{ci}}$ is the impurity concentration, $k_{\text{F}}=E_{\text{F}}/(\hbar v)$ is the Fermi momentum and $\alpha_{\text{g}}=e^2/(\hbar v\tilde{\kappa})\approx 1.2$ is the effective graphene fine-structure constant.

One can see that multiple parameters affect $\hbar/\tau_{\text{ci}}$, contributing rather differently to its strain dependence. First, the density of states $g(E_{\text{F}})$, which is now in the denominator, tends to decrease the scattering rate, in contrast to the case of point defects. Second, applying strain increases, peculiarly, the graphene fine-structure constant via a reduced Fermi velocity: $\alpha_{\text{g}}/\alpha_{\text{g},0}\approx 1 - \nu\epsilon$. From Eq. (\ref{ScatterCI}) it follows that increasing $\alpha_{\text{g}}$ leads to the growth of the scattering rate. Third, the scattering rate is affected by the graphene-impurity distance and by the Fermi momentum via their dimensionless product $k_{\text{F}}z$. As the Fermi energy is fixed, the Fermi momentum increases with strain: $k_{\text{F}}/k_{\text{F},0}\approx 1 - \nu\epsilon$, which reduces scattering. However, if $z$ decreases with strain, this may potentially increase of $\hbar/\tau_{\text{ci}}$. Since we apply strain by bending the PET substrate, it is possible that this brings the graphene closer to the substrate  (where we assume the charged scatterers are located). If $z\sim 1-2$ nm then $k_{\text{F}}z\sim$ 1, which a regime where the scattering is highly sensitive to the graphene-substrate separation \cite{HwangPRL07}.

By combining these competing factors, we obtain:
\begin{equation}\label{ScatterPDLinear}
c_{\text{ci}}\approx -1 + 2\nu - \frac{\partial \ln I} {\partial \ln \alpha_{\text{g}}} \nu + \frac{\partial \ln I} {\partial \ln (k_{\text{F}}z)}\left(-\nu + \zeta\right),
\end{equation}
where $I$ is the integral in the Eq. (\ref{ScatterCI}) and $\zeta=\partial\ln z/\partial \epsilon$. Specifically, assuming that $k_{\text{F}}z=1$, we numerically obtain $c_{\text{ci}} \approx -3.5 - 2.1\zeta$.  One can see that the total contribution of other factors than the distance $z$ would result in the opposite strain dependence as compared to the experiment. Thus only the change of $z$ may account for the effect that we observe. Specifically, if $\zeta < -8$ , \emph{i.e. }$z$ decreases by more than 8\% at 1\% of applied strain, then $c_{\text{ci}} > 13.4$, which would quantitatively explain our experimental observation. Although we cannot measure $z$ and $\zeta$ directly, these estimates do not seem to be unreasonable, given that CVD graphene is bound very weakly to the substrate. Consequently, we speculate that the increase of scattering by charged impurities due to the reduction of the effective graphene-PET distance might be an explanation of our experimental observation.

\emph{Acoustic phonons.} Scattering from the acoustic phonons was calculated in \cite{StauberPRB07,HwangPRB08}:
\begin{equation}\label{ScatterPh}
\frac{\hbar}{\tau_{\text{ap}}} = \frac{\pi}{4} g(E_{\text{F}})\left(\frac{k_{B}T}{\rho v_{\text{ap}}^2}\right)F^2,
\end{equation}
\noindent where $\rho$ is the mass density, $v_{\text{ap}}$ is the sound velocity (averaged over the longitudinal and transverse branches),  $F$ is the deformation potential, and $k_{B}$ is the Boltzmann constant.
When strain is applied, both the mass density and the sound velocity decrease: $\rho/\rho_{0}\approx 1-\epsilon$ and $v_{\text{ap}}/v_{\text{ap},0}\approx 1-h\epsilon$. Based on theoretical results of  Ref.\cite{BaimovaPSS12}, we can estimate that $h\approx 1.4$. Ignoring a possible strain-induced decrease of the deformation potential, we obtain:
\begin{equation}\label{ScatterPD}
c_{\text{ap}} \approx -2\nu + 1 + h \approx 3.7.
\end{equation}

\noindent Since $c_{\text{ap}}$ is well below the experimental value of 13.4, one can also exclude the acoustic phonons as a dominant contribution to the strain-induced increase of the total scattering rate.

\emph{Surface phonons. } At room temperature, surface optical phonons in polar substrates such as SiO$_{2}$ and SiC also contribute to scattering in the overlying graphene \cite{HwangPRB13}. We are not aware whether or not this mechanism is at work for a polymer substrate like ours. The analytical formula for $\hbar/\tau_{\text{sp}}$ is too complicated to be presented here. Importantly, however, the dynamical screening gives rise to a strong dependence of the surface-phonon scattering on the distance $z$, similar to the case of charge impurities. Thus, the effect of the surface phonons cannot be excluded at the moment as a candidate for the strain-induced increase of $\hbar/\tau$.

\emph{Grain boundaries.} In CVD graphene, grain boundaries are known  to diminish the effective transport mobility \cite{SongSR12}. The boundary region acts as a resistance in series with the bulk graphene, which can be expressed in terms of an additional scattering rate for the static conductivity. In optics, however, the situation is different. As the boundary regions represent a small fraction of the total graphene area, the effect of the boundaries is to excite graphene plasmons due to the breaking of translational symmetry, rather than to increase the Drude scattering rate. This would shift the Drude peak to finite energies \cite{CrasseeNL12}, which would produce a totally different spectral shape of the differential conductivity than what we observe in \fref{FigCond}(c). Thus, we can exclude the grain boundaries from the candidate list.

\subsection{Effect of strain on the charge mobility and electromagnetic absorption}

Based on the observed growth of the scattering rate we can estimate its effect on other important parameters, such as the mobility and the terahertz/microwave absorption. If the Fermi level is fixed, the anisotropic mobility is determined by the anisotropic Fermi velocity and the scattering rate (which we assume to be isotropic):
\begin{equation}\label{MuAniso}
\mu_{\parallel, \perp} = \frac{e^2\tau v_{\parallel, \perp}^2}{E_{\text{F}}},
\end{equation}
\noindent from which it follows that
\begin{equation}\label{MuLinear}
\frac{\mu_{\parallel, \perp}}{\mu_{0}} \approx 1 + (2\nu_{\parallel,\perp}-c)\epsilon.
\end{equation}

\noindent Using the experimentally obtained value of $c$ and theoretical values for $\nu_{\parallel,\perp}$ we find that $\mu_{\parallel}$ and $\mu_{\perp}$ should decrease with a rate of 17\% and 12.5\% per 1\% of strain, respectively.

The (anisotropic) optical absorption in the low-THz and the microwave range, where $\omega\tau \ll 1$, is proportional to the zero-frequency limit of the (anisotropic) optical conductivity:
\begin{equation}\label{SigmaAniso}
\sigma_{\text{dc},\parallel, \perp}=\frac{D_{\parallel, \perp}\tau}{\pi},
\end{equation}
\noindent which yields in the linear approximation:
\begin{equation}\label{SigmaLinear}
\frac{\sigma_{\text{dc},\parallel, \perp}}{\sigma_{\text{dc},0}}\approx 1 + (\pm \nu_{\parallel}\mp\nu_{\perp}-c)\epsilon.
\end{equation}
\noindent Therefore $\sigma_{\text{dc},\parallel}$ and $\sigma_{\text{dc},\perp}$, and the corresponding low-frequency absorption coefficients should decrease with a rate of 15.7\% and 11.1\% per 1\% of strain, respectively.

Although the strain value in our experiment was limited to about 0.8\%, graphene can in principle support strains of up to 25\% \cite{LiuPRB07,KimNature09} depending on the substrate used. The large numbers discussed above therefore suggest the possibility to controllably and reproducibly vary the mobility and optical absorption of graphene by several tens of percent, even if the linear dependence observed in our present study will eventually saturate.

\section{Conclusion}

We investigated the far-infrared optical conductivity of CVD-graphene on a flexible PET substrate under a tensile uniaxial strain, nominally up to 2\%. These experiments were supported by AFM/KPFM measurements that revealed a flattening of wrinkles and folds initially present in graphene. Using Raman spectroscopy allowed us to calibrate the effective strain values, which was found to be smaller than the nominal strain, probably due to this wrinkle relaxation. Fitting the far-infrared spectra to the Drude model allowed us to accurately determine the Drude weight and the optical scattering rate independently as functions of the applied strain.

We find that the Drude weight remains essentially constant, implying that the Fermi level is quenched at a fixed level, probably by the presence of the substrate acting as a charge reservoir to compensate the strain-induced changes in the density of states of graphene. Additionally, we observe a strong linear increase of the scattering rate (by about 13\% per 1\% of the applied strain). This effect is found to be reversible and reproducible during strain cycling. A detailed theoretical analysis suggests that a likely origin of this unexpected effect is a change of the distance between graphene and the substrate, which influences, via a modified dynamical screening, the electronic scattering by the charged impurities or polar surface phonons in the substrate. In both cases, this mechanism would be truly unique for a 2D material, where the substrate plays a critical role.

Regardless of the actual physical mechanism, the strong effect of strain on the scattering rate may have important implications for graphene-based flexible optoelectronic devices, as it affects the electronic mobility and low-energy electromagnetic absorption. In particular, CVD graphene on plastic substrates is becoming increasingly important for flexible photodetectors, touch screens and microelectromechanical systems (MEMS). Our observation that these properties can be controlled mechanically, is therefore important for benchmarking and diversifying the functionalities of such devices.

\section{Acknowledgements}

This research was supported  by the EU Graphene Flagship (Contract No. CNECT-ICT-604391 and 696656) and by the Swiss National Science Foundation (Grants No. 200020-156615 and 200021-15317).  PP and IG acknowledge UniGE COINF support allowing the development of the drift correction differential analysis algorithm. The authors are grateful to S. Muller, M. Brandt, J. Teyssier, J.-M. Poumirol and P. T\"uckmantel for discussions and technical assistance.



\begin{thebibliography}{99}
\bibitem{BonaccorsoNP10} Bonaccorso F, Sun Z, Hasan T and Ferrari A C 2010 Graphene photonics and optoelectronics \textit{Nat. Photon.} \textbf{4} 611
\bibitem{JangAM16} Jang H, Park Y J, Chen X, Das T, Kim M S and Ahn J H 2016 Graphene-Based Flexible and Stretchable Electronics \textit{Adv. Mater.} \textbf{28} 4184
\bibitem{LeeScience08} Lee C, Wei X, Kysar J W and Hone J 2008 Measurement of the Elastic Properties and Intrinsic Strength of Monolayer Graphene \textit{Science} \textbf{321} 385
\bibitem{LiuPRB07} Liu F, Ming P and Li J 2007 \textit{Ab initio} calculation of ideal strength and phonon instability of graphene under tension \textit{Phys. Rev. B} \textbf{76} 064120
\bibitem{KimNature09} Kim K S \etal~ 2009 Large-scale pattern growth of graphene films for stretchable transparent electrodes \textit{Nature} \textbf{457} 706
\bibitem{ShioyaNL15} Shioya H, Russo S, Yamamoto M, Craciun M F and Tarucha S 2015 Electron states of Uniaxially Strained Graphene \textit{Nano Lett.} \textbf{15} 7943
\bibitem{HeAPL15} He X \etal~ 2015 Tuning the graphene work function by uniaxial strain \textit{Appl. Phys. Lett.} \textbf{106} 043106
\bibitem{YuJPCCL08} Yu T, Ni Z, Du C, You Y, Wang Y and Shen Z 2008 Raman Mapping Investigation of Graphene on Transparent Flexible Substrate: The Strain Effect \textit{J. Phys. Chem. Lett.} \textbf{112} 12602
\bibitem{MohiuddinPRB09} Mohiuddin T M G \etal~ 2009 Uniaxial strain in graphene by Raman spectroscopy: G peak splitting, Gruneisen parameters and sample orientation \textit{Phys. Rev. B} \textbf{79} 205433
\bibitem{CorroJPC15} Corro E D, Kavan L, Kalbac M and Frank O 2015 Strain Assessment in Graphene through the Raman 2D mode \textit{J. Phys. Chem. C} \textbf{119} 25651
\bibitem{NiAM14} Ni G X, Yang H Z, Ji W, Baeck S J, Toh C T, Ahn J H, Pereira V M and Ozyilmaz B 2014 Tuning Optical Conductivity of Large-Scale CVD Graphene by Strain Engineering \textit{Adv. Mater.} \textbf{26} 1081
\bibitem{PellegrinoHPR10} Pellegrino F M D, Angilella G G N and Pucci R 2010 Effect of uniaxial strain on the Drude weight of graphene \textit{High Pressure Res.} \textbf{31} 98
\bibitem{PereiraPRB09} Pereira V M,  Castro Neto A H and Peres N M R 2009 Tight-binding approach to uniaxial strain in graphene \textit{Phys. Rev. B} \textbf{80} 045401
\bibitem{RibeiroNJP09} Ribeiro R M, Pereira V M, Peres N M R, Briddon P R and Castro Neto A H 2009 Strained graphene: tight-binding and density functional calculations \textit{New J. Phys.} \textbf{11} 115002
\bibitem{KimAPL12} Kim J Y, Lee C, Bae S, Kim S J, Kim K S, Hong B H and Choi E J 2012 Effect of uni-axial strain on THz/far-infrared response of graphene \textit{Appl. Phys. Lett.} \textbf{100} 041910
\bibitem{Gaponenko_ms} Gaponenko I, T{\"u}ckmantel P, Ziegler B, Rapin G, Chhikara M and Paruch P Computer vision distortion correction of scanning probe microscopy images, submitted to \textit{Sci. Rep.}
\bibitem{KuzmenkoRSI05} Kuzmenko A B 2005 Kramers–Kronig constrained variational analysis of optical spectra \textit{Rev. Sci. Instrum.} \textbf{76} 083108
\bibitem{HorngPRB11} Horng J \etal~ 2011 Drude conductivity of Dirac fermions in graphene \textit{Phys. Rev. B} \textbf{83} 165113
\bibitem{SensaleNatComm12} Sensale-Rodriguez B, Yan R, Kelly M M, Fang T, Tahy K, Hwang W S, Jena D, Liu L and Xing H G 2012 Broadband graphene terahertz modulators enabled by intraband transitions \textit{Nat. Commun.} \textbf{3} 780
\bibitem{AndoJPSJ02} Ando T,Zheng Y and Suzuura H 2002 Dynamical Conductivity and Zero-Mode Anomaly in Honeycomb Lattices \textit{J. Phys. Soc. Jpn.} \textbf{71} 1318
\bibitem{RenNL12} Ren L \etal~ 2012 Terahertz and Infrared Spectroscopy of Gated Large-Area Graphene \textit{Nano Lett.} \textbf{12} 3711
\bibitem{KuzmenkoPRL08} Kuzmenko A B, van Heumen E, Carbone F and van der Marel D 2008 Universal Optical Conductance of Graphite \textit{Phys. Rev. Lett.} \textbf{100} 117401
\bibitem{NairScience08} Nair R R, Blake P, Grigorenko A N, Novoselov K S, Booth T J, Stauber T, Peres N M R and Geim A K \etal 2008 Fine Structure Constant Defines Visual Transparency of Graphene \textit{Science} \textbf{320} 1308
\bibitem{PellegrinoPRB10} Pellegrino F M D, Angilella G G N and Pucci R 2010 Strain effect on the optical conductivity of graphene \textit{Phys. Rev. B} \textbf{81} 035411
\bibitem{PereiraEPL10} Pereira V M, Ribeiro R M, Peres N M R and Castro Neto A H 2010 Optical properties of strained graphene \textit{EPL} \textbf{92} 67001
\bibitem{BaimovaPSS12} Baimova Y A, Dmitriev S V, Savin A V, and Kivshar Y S 2012 Velocities of sound and the densities of phonon states in a uniformly strained flat graphene Sheet \textit{Phys. Solid State} \textbf{54} 866
\bibitem{ShahMPLB13} Shah R, Mohiuddin T M G and Singh R N 2013 Giant Reduction of Charge Mobility in Strained Graphene \textit{Mod. Phys. Lett. B} \textbf{27} 1350021
\bibitem{AndoJPSJ06} Ando T 2006 Screening Effect and Impurity Scattering in Monolayer Graphene \textit{J. Phys. Soc. Jpn.} \textbf{75} 074716
\bibitem{TanPRL07} Tan Y W, Zhang Y, Bolotin K, Zhao Y, Adam S, Hwang E H, Das Sarma S, Stormer H L and Kim P \etal 2007 Measurement of Scattering Rate and Minimum Conductivity in Graphene, Phys. Rev. Lett. \textbf{99} 246803
\bibitem{HwangPRL07} Hwang E H, Adam S, and Das Sarma S 2007 Carrier Transport in Two-Dimensional Graphene Layers \textit{Phys. Rev. Lett.} \textbf{98} 186806
\bibitem{StauberPRB07} Stauber T, Peres N M R and Guinea F 2007 Electronic transport in graphene: A semiclassical approach including midgap states \textit{Phys. Rev. B} \textbf{76} 205423
\bibitem{ChenNP08}Chen J H, Jang C, Adam S, Fuhrer M S, Williams E D and Ishigami M 2008 Charged-impurity scattering in graphene \textit{Nat. Phys.} \textbf{4} 377
\bibitem{HwangPRB08} Hwang E H and Das Sarma S 2008 Acoustic phonon scattering limited carrier mobility in two-dimensional extrinsic graphene \textit{Phys. Rev. B} \textbf{77} 115449
\bibitem{PonomarenkoPRL09} Ponomarenko L A \etal 2009 Effect of a High-k Environment on Charge Carrier Mobility in Graphene \textit{Phys. Rev. Lett.} \textbf{102} 206603
\bibitem{SongSR12} Song H S, Li S L, Miyazaki H, Sato S, Hayashi K, Yamada A, Yokoyama N and Tsukagoshi K 2012 Origin of the relatively low transport mobility of graphene grown through chemical vapor deposition \textit{Sci. Rep.} \textbf{2} 337
\bibitem{HwangPRB13} Hwang E H and Das Sarma S 2013 Surface polar optical phonon interaction induced many-body effects and hot-electron relaxation in graphene \textit{Phys. Rev. B} \textbf{87} 115432
\bibitem{YuPRB16} Yu K, Kim J, Kim J Y, Lee W, Hwang J Y, Hwang E H and Choi E J \etal 2016 Infrared spectroscopic study of carrier scattering in gated CVD graphene \textit{Phys. Rev. B} \textbf{94} 235404
\bibitem{BlaksleeJAP70} Blakslee O L, Proctor D G, Seldin E J, Spence G B and Weng T 1970 Elastic Constants of Compression-Annealed Pyrolytic Graphite \textit{J. Appl. Phys.}  \textbf{41} 3373


\bibitem{CrasseeNL12} Crassee I, Orlita M, Potemski M, Walter A L, Ostler M, Seyller Th, Gaponenko I, Chen I and Kuzmenko AB 2012 Intrinsic Terahertz Plasmons and Magnetoplasmons in Large Scale Monolayer Graphene \textit{Nano Lett.} \textbf{12} 2470
\end{thebibliography}

\end{document}


\title{Supplementary Information: Effect of uniaxial strain on the optical Drude scattering in graphene}

\author{Manisha Chhikara}
\affiliation{Department of Quantum Matter Physics, University of Geneva, CH-1211 Geneva 4, Switzerland}
\author{Iaroslav Gaponenko}
\affiliation{Department of Quantum Matter Physics, University of Geneva, CH-1211 Geneva 4, Switzerland}
\author{Patrycja Paruch}
\affiliation{Department of Quantum Matter Physics, University of Geneva, CH-1211 Geneva 4, Switzerland}
\author{Alexey. B. Kuzmenko}
\email{Alexey.Kuzmenko@unige.ch}, \email{Manisha.Chhikara@unige.ch}
\affiliation{Department of Quantum Matter Physics, University of Geneva, CH-1211 Geneva 4, Switzerland}

\date{\today}
\maketitle

\section{Strain-dependent far-infrared transmission spectra of the PET substrate}

\begin{figure*}
\includegraphics[width=17cm]{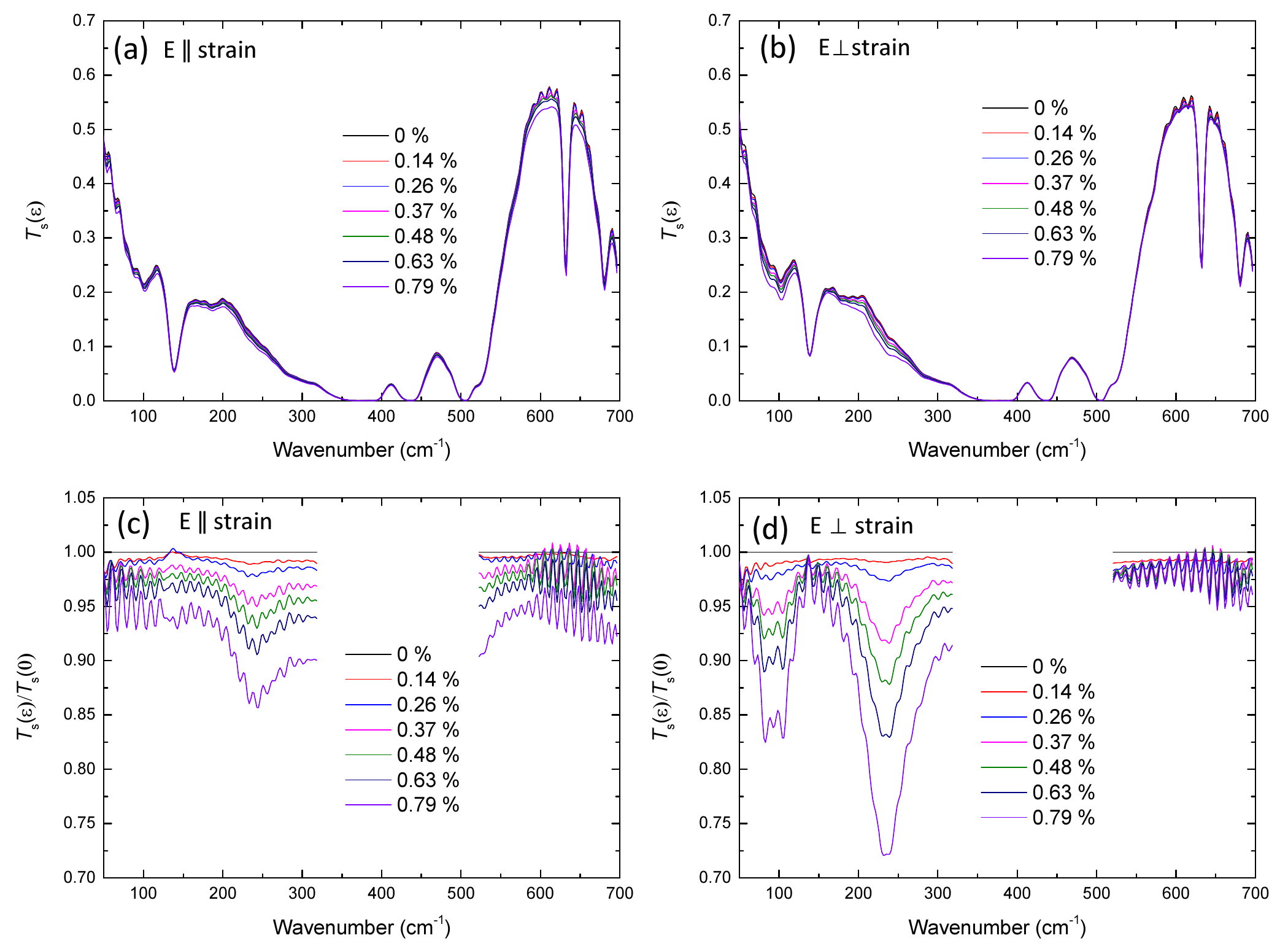}
\caption{(a) and (b) Absolute transmission of a reference PET substrate (250 $\mu$m thick) at different bending radii corresponding to different applied strain values $\epsilon$. (a) and (b) correspond to the polarization parallel and perpendicular to the strain axis, respectively. (c) and (d) represent the same spectra normalized to zero strain in order to emphasize the photoelastic effects.}
\label{FigSubst}
\end{figure*}

In \fref{FigSubst}(a) and (b) we present the absolute far-infrared transmission spectra $T_{\text{s}}(\epsilon,\omega)$ of a reference polyethylene terephthalate (PET) substrate (thickness 250 $\mu$m) at different bending radii corresponding to different applied uniaxial strain values $\epsilon$ for the polarization of light parallel and perpendicular to the strain axis, respectively. The spectra, which are dominated by optical phonon absorption, are similar for the two directions, as expected for an isotropic material.

\begin{figure*}
\includegraphics[width=8cm]{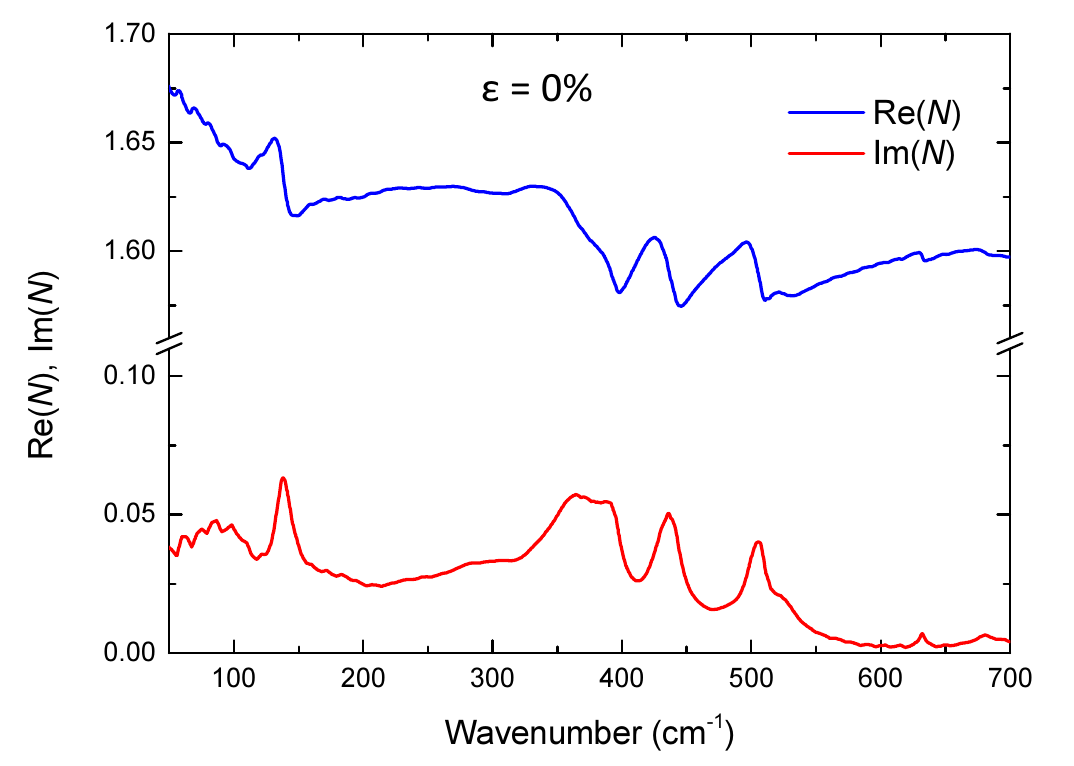}
\caption{The complex refractive index of PET at zero strain (parallel polarization), extracted from the transmission using a Kramers-Kronig analysis.}
\label{FigSubstN1N2}
\end{figure*}

At each strain value, we extracted the substrate complex dielectric function $\epsilon_{\text{PET}}(\epsilon,\omega)$, and consequently the complex refractive index $N(\epsilon,\omega)=\sqrt{\epsilon_{\text{PET}}(\epsilon,\omega)}$ by applying a Kramers-Kronig constrained variational analysis \cite{KuzmenkoRSI05} to the measured transmission spectra. An example of the real and the imaginary parts of $N$ at zero strain is shown in \fref{FigSubstN1N2}. One can see that $\text{Re}(N)$ varies weakly in the interval 1.58-1.67 and $\text{Im}(N) \ll\text{Re}(N)$ everywhere in the considered spectra range.

In order to see better the strain/bending induced photoelastic effects, in \fref{FigSubst}(c) and (d) we plot the zero strain normalized spectra $T_{\text{s}}(\epsilon,\omega)/T_{\text{s}}(0,\omega)$. One can see that the photoelastic effects, unlike the absolute transmission, demonstrate a marked anisotropy with respect to the strain axis. The normalized spectra show pronounced dips at 230 cm$^{-1}$ for the parallel polarization, and at 100 and 230 cm$^{-1}$ for the perpendicular one. These frequencies do not coincide with any of the optical phonons seen in the absolute spectra at zero strain. Therefore, these features are likely to be signatures of extra modes optically activated by the deformation. With a good accuracy, the photoelastic spectra for both polarizations at all strain values can be described by the relation:

\begin{equation}\label{PhoelRatio}
\frac{T_{\text{s}}(\epsilon,\omega)}{T_{\text{s}}(0,\omega)}\approx 1 + P(\omega)\epsilon,
\end{equation}
\noindent where $P(\omega)$ is a generic (strain independent) photoelastic function.

\section{Correction procedure for a sample-reference strain mismatch}

The presence of strong photoelastic effects in the substrate makes the determination of the extinction spectra of graphene $1-T_{\text{g}}(\epsilon,\omega)/T_{\text{s}}(\epsilon,\omega)$ a delicate procedure. Even though we took extreme care to match the strain value in the sample and the reference measurements, certain substrate-related artefacts sometimes appear in the directly extracted extinction spectra. Therefore we established a robust correction procedure, which relies on the fact that the photoelastic function $P(\omega)$ has very strong characteristic spectral features, which are totally uncorrelated with the Drude absorption in graphene.

First, we carefully measured the function $P(\omega)$. Then we fitted the experimental extinction spectra using the modified Eq. (1) from the main text:
\begin{equation}\label{PhoelRatio}
1-\frac{T_{\text{g}}}{T_{\text{s}}}=1-\frac{1 + P\delta \epsilon}{\left| 1+f_{\text{s}} Z_{0}\sigma/2\right|^{2}},
\end{equation}
\noindent where the optical conductivity of graphene was calculated using the Drude formula (Eq. (3) from the main text), while the strain mismatch $\delta \epsilon \ll \epsilon$ was treated as a third adjustable parameter. Due to the sharpness of the spectral dips in $P(\omega)$, such a fit provides a reliable value of $\delta\epsilon$ independently from the Drude parameters $D$ and $\tau$. This value was finally used to calculate the corrected extinction spectra:
\begin{equation}\label{ExtCorr}
1-\left(\frac{T_{\text{g}}}{T_{\text{s}}}\right)_{\text{corr}}=1-\frac{1}{1 + P\delta \epsilon}\left(\frac{T_{\text{g}}}{T_{\text{s}}}\right)_{\text{meas}}.
\end{equation}

\section{Anisotropy of the Fermi velocity in uniaxially strained graphene}

The Fermi velocity in uniaxially strained graphene within the tight-binding model was calculated in Refs.[\onlinecite{PellegrinoPRB10}] and [\onlinecite{PereiraEPL10}]. Considering the case where the hopping elements $t_{1}$, $t_{2} $ and $t_{3}$ change when the strain is applied, but ignoring the modification of the lattice constants, the Fermi velocity components parallel and perpendicular to the strain axis are the following:
\begin{equation}\label{Vfermi}
\frac{v_{\parallel,\perp}}{v_{0}}=\frac{1}{\sqrt{3}t_{0}}\left\{t_{1}^2+t_{2}^2+t_{3}^2\mp 2\left(t_{1}^4+t_{2}^4+t_{3}^4-t_{1}^2t_{2}^2-t_{2}^2t_{3}^2-t_{1}^2t_{3}^2\right)^{1/2}\right\}^{1/2},
\end{equation}
\noindent where $v_{0}=3a_{0}t_{0}/2\hbar$ is the Fermi velocity in unstrained graphene.

For small strain values $\epsilon$,  a linear expansion can be used for the hopping parameters:
\begin{equation}\label{HoppingLinear}
\frac{t_{1}}{t_{0}}\approx 1 + \xi_{1}\epsilon, \frac{t_{2}}{t_{0}}\approx 1 + \xi_{2}\epsilon, \frac{t_{3}}{t_{0}}\approx 1 + \xi_{3}\epsilon
\end{equation}
\noindent Applying these linearized formulas to Eq. (\ref{Vfermi}), we obtain:
\begin{equation}\label{VfermiLinear}
\frac{v_{\parallel,\perp}}{v_{0}}\approx 1 + \nu_{\parallel,\perp}\epsilon
\end{equation}
\noindent with
\begin{equation}\label{Nu}
\nu_{\parallel,\perp}=\frac{2}{3}\left(\xi_{1}+\xi_{2}+\xi_{3}\right)\mp\frac{4}{3}\left(\xi_{1}^{2}+\xi_{2}^{2}+\xi_{3}^{2}-\xi_{1}\xi_{2}-\xi_{2}\xi_{3}-\xi_{1}\xi_{3}\right)^{1/2}.
\end{equation}

The dependence of the hopping parameters was calculated \cite{RibeiroNJP09} for two orthogonal strain directions with respect to the crystal structure using an \emph{ab-initio} approach. Based on their results (Figure 3 from Ref.[\onlinecite{RibeiroNJP09}]), we can estimate that for the zig-zag direction $\xi_{1}=\xi_{3}\approx -2.0$ and $\xi_{2}\approx 0.5$, while for the armchair direction $\xi_{1}=\xi_{3}\approx -0.4$ and $\xi_{2}\approx 2.8$. Interestingly, in both cases Eq. (\ref{Nu}) yields: $\nu_{\parallel}\approx-2.8$, $\nu_{\perp}\approx 0.5$, which means that these coefficients are almost insensitive to the strain orientation with respect to the crystal lattice.

Although the authors of Refs.[\onlinecite{PellegrinoPRB10}] and [\onlinecite{PereiraEPL10}] did not consider the change of the lattice constants and the Brillouin zone, for the present purpose one has to take into account that the lattice stretches parallel to strain, while it remains the same along the perpendicular direction (ignoring a small Poisson ratio). As the Fermi velocity at fixed values of the hopping parameters scales linearly with the lattice constant, including this effect increases the linear-expansion coefficient $\nu_{\parallel}$ exactly by 1. Thus, in this paper we use the theoretical values $\nu_{\parallel}\approx-1.8$ and $\nu_{\perp}\approx +0.5$.

\section{Scattering by charged impurities including dynamical screening}

Within the relaxation time approximation, the scattering rate by charged impurities (ci) is calculated as follows \cite{AndoJPSJ06,HwangPRL07,StauberPRB07}:
\begin{equation}\label{Scatter}
\frac{\hbar}{\tau_{\text{ci}}}=n_{\text{ci}}\ g(E_{\text{F}})\int_{0}^{\pi} d\theta (1-\cos^2\theta)\left|\frac{V(q)}{\epsilon(q)}\right|^2,
\end{equation}
\noindent where $n_{\text{ci}}$ is the two-dimensional concentration of scatterers, $g(E_{\text{F}})=2E_{\text{F}}/(\pi\hbar^{2}v^{2})$ is the density of states at the Fermi level $E_{\text{F}}$, $q=2k_{\text{F}}\sin(\theta/2)$, $v$ is the Fermi velocity,  $k_{\text{F}}=E_{\text{F}}/(\hbar v)$, $V(q)$ is the effective momentum dependent Coulomb potential and $\epsilon(q)$ is the effective dielectric constant.

For the Coulomb potential, we have\cite{HwangPRL07}:
\begin{equation}\label{Vq}
V(q)=\frac{2\pi e^2}{\tilde{\kappa}q}e^{-qz}
\end{equation}
\noindent where $\tilde{\kappa}$ is the average dielectric function of the medium surrounding graphene and $z$ is the distance between graphene and the plane containing the charged scatterers. Within the random-phase approximation, the dielectric function is calculated as follows \cite{AndoJPSJ06,HwangPRL07}:
\begin{equation}\label{Epsilonq}
\epsilon(q)=1 +  \frac{4e^2k_{\text{F}}}{\hbar\tilde{\kappa}vq}\left(1-\frac{\pi q}{8 k_{\text{F}}}\right)= 1 +  \alpha_{\text{g}}\left(\frac{4k_{\text{F}}}{q}-\frac{\pi}{2}\right),
\end{equation}
\noindent where $\alpha_{\text{g}}=e^2/(\hbar\tilde{\kappa}v)$ is the effective graphene fine-structure constant. By combining the above equations we arrive at the formula used in the main text:

\begin{equation}\label{ScatterShort}
\frac{\hbar}{\tau_{\text{ci}}} = \frac{n_{\text{ci}}}{g(E_{\text{F}})}\int_{0}^{\pi}\frac{e^{-4k_{\text{F}}z\sin\frac{\theta}{2}}\sin^2\theta d\theta}{\left[1-\left(\frac{\pi}{4}-\frac{1}{2\alpha_{\text{g}}}\right)\sin\frac{\theta}{2}\right]^2}.
\end{equation}